\documentclass[twocolumn,prb,10pt,aps,floatfix]{revtex4-1}
\usepackage[english]{babel}
\usepackage{graphicx}
\usepackage{amsmath}
\usepackage[colorlinks,linkcolor=blue,citecolor=blue,urlcolor=blue]{hyperref}
\usepackage{color}
\usepackage{soul}
\begin{document}
\title{Quasiballistic transport in long-range anisotropic Heisenberg model}

\author{M. Mierzejewski}
\author{J. Wronowicz}
\author{J. Pawłowski}
\author{J. Herbrych}
\affiliation{Department of Theoretical Physics, Faculty of Fundamental Problems of Technology, Wroc{\l}aw University of Science and Technology, 50-370 Wroc{\l}aw, Poland}
\date{\today}
\begin{abstract}
Purely ballistic transport is a rare feature even for integrable models. By numerically studying the Heisenberg chain with the power-law exchange, \mbox{$J\propto1/r^\alpha$}, where $r$ is a distance, we show that for spin anisotropy $\Delta \simeq \exp(-\alpha+2)$ the system exhibits a quasiballistic spin transport and the presence of fermionic excitation which do not decay up to extremely long times $\sim10^3/J$. This conclusion is reached on the base of the dynamics of spin domains, the dynamical spin conductivity, inspecting the matrix elements of the spin-current operator, and by the analysis of most conserved operators. Our results smoothly connects two models where fully ballistic transport is present: free particles with nearest-neighbor hopping and the isotropic Haldane-Shastry model.
\end{abstract}
\maketitle

\section{Introduction}

Range and type of interaction play the most crucial role in determining the properties of quantum many-body systems. Phenomena like ballistic propagation in integrable one-dimensional (1D) Heisenberg model \cite{Zotos1997,Bertini2021}, exotic magnetism due to frustration \cite{Nisoli2017}, unusual phase transitions \cite{Sandvik2010,Yang2021}, highly entangled spin liquids \cite{Balents2010}, to name a few, essentially depend on the type of exchange present in the system. Nowadays, the experimental progress in quantum simulators, i.e., cold-atoms in optical lattice \cite{Bloch2006,Hild2014,Gross2017,Zhang2017,Schafer2020,Jepsen2020,Wei2022}, also with Rydberg states \cite{Weimer2010,Anderson2011,Labuhn2016,Bernien2017,Sanchez2018,Browaeys2020,Borish2020,Geier2021,Hollerith2022,Scholl2022,Steinert2022}, allow for the study of such systems with unprecedented precision. Furthermore, it allows also for the discovery of new phenomena. Especially the systems with direct long-range exchange (much beyond the nearest-neighbor spacing) are of great interest since they appear only as an effective low-energy description in the solid-state setups. For example, the transverse Ising model with power-law decaying long-range interaction was successfully created \cite{Britton2012,Joshi2022} and studied \cite{Richerme2014,Jurcevic2014} in the context of the Lieb-Robinson bound on the information propagation in the system.

Although in recent years, there was a lot of work devoted to the nonlocality in the long-range models \cite{Hauke2013,Feug2015,Maghrebi2016,Lepori2017,Frerot2017,Vanderstraeten2018,Cevolani2018,Kloss2019,Lerose2019,Ren2020,Schneider2021,Birnkammer2022,Bulchandani2022}, the dynamical properties of such systems are mostly unknown. Since the latter are often directly probed in the experimental setups (e.g., in the density expansion experiments), in this work we discuss the spin transport in the anisotropic Heisenberg chain with long-range exchange, $J(r)=J/r^{\alpha}$. Studying the dynamics of spin domains (Sec.~\ref{sec1}), the dynamical spin conductivity (Sec.~\ref{sec2}), inspecting the matrix elements of the spin-current operator (Sec.~\ref{sec4}), and analyzing the most conserved current operators (Sec.~\ref{sec5}), we show that the model exhibits a transient ballistic transport persisting up to very long times $t/J \sim 10^2-10^3$. Such quasiballistic behaviour is shown to exists along a sharp line in the space of model parameters that smoothly interpolates between two purely ballistic models: free particles with nearest-neighbor hopping for $\alpha \to \infty$ and the isotropic Haldane-Shastry model for $\alpha=2$. The former is {\it trivially} exactly solvable \cite{Lieb1961} via the Fourier transform to the momentum space. The latter was originally introduced independently by Haldane \cite{Haldane1988} and Shastry \cite{Shastry1988} as a solution of the 1D resonating-valence-bond state \cite{Anderson1973}. Interestingly, not only these models are integrable, but also the spin current, $j$, is a constant of motion of these Hamiltonians, $[H,j]=0$ in the thermodynamic limit \cite{Essler1995,Greiter2005,Sechin2018}. One should contrast this behaviour with the nearest-neighbor anisotropic Heisenberg model (nn-AHM) which is integrable but $[H,j]\ne0$.

In the following, we study the 1D \mbox{spin-$1/2$} anisotropic Heisenberg model with long-range exchange (long-range AHM),
\begin{equation}
 H=\sum_{\ell,r} J(r)\left[ \frac{1}{2}\left(S^{+}_{\ell}S^{-}_{\ell+r}
 +S^{-}_{\ell}S^{+}_{\ell+r}\right)+\Delta S^{z}_{\ell}S^{z}_{\ell+r}\right]\,,
\label{ham}
\end{equation}
with $\Delta$ as an anisotropy in $z$-direction and $J(r)=J/r^\alpha$ as a long-range power-law spin exchange, where $J=1$ sets the unit of energy (as well as $\hbar=k_B=1$) and $\alpha>1$ controls the decay of exchange. The $\alpha\to\infty$ limit recovers the nn-AHM, which is integrable and can be mapped via the Jordan-Wigner transformation\cite{zvyagin} to spinless fermions interacting with the strength $\Delta$. Note that for any finite $\alpha$, this transformation maps the spin chain on interacting fermions also for $\Delta=0$.

\begin{figure*}[!t]
\includegraphics[width=1.0\textwidth]{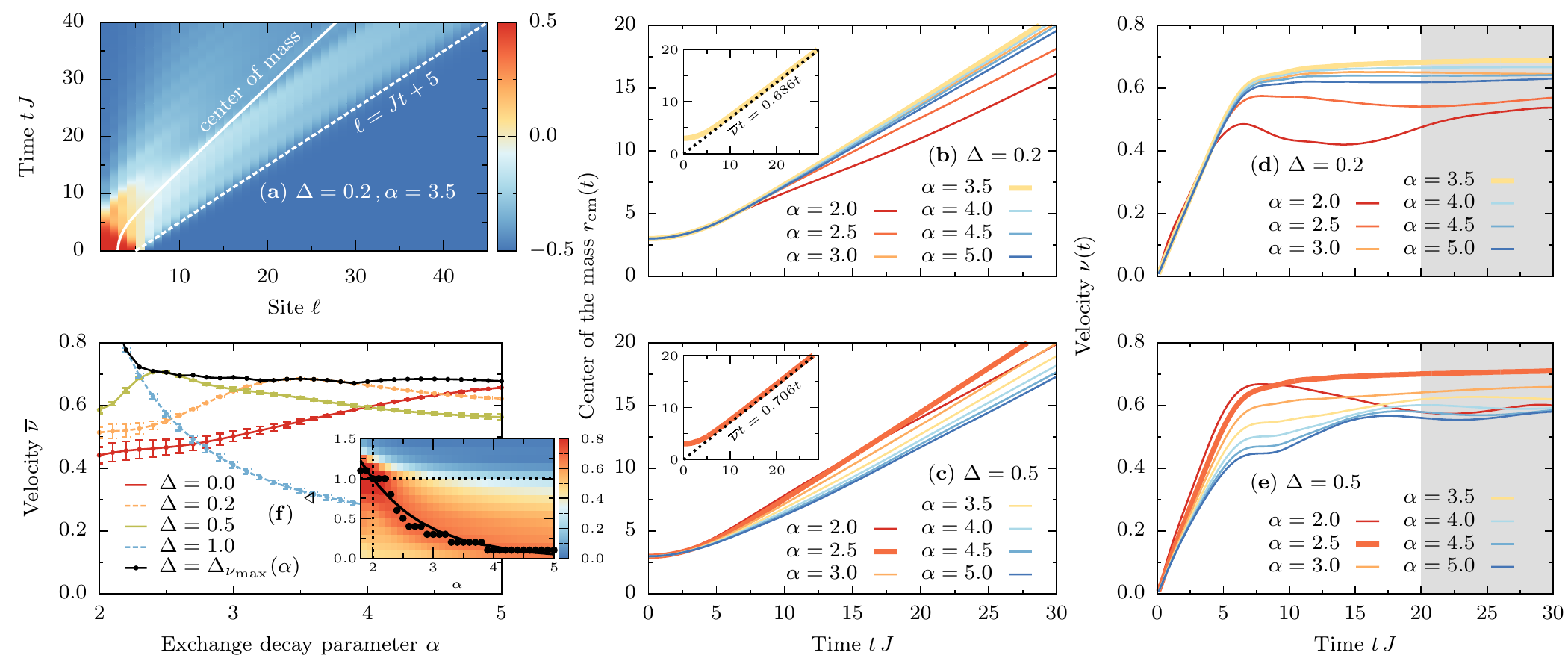}
\caption{({\bf a}) Time-dependence of the spin density, $\langle S^z_\ell\rangle$, for the expansion from the nearly down-polarized state with $5$ up-spins at the boundary of a $L=45$ sites chain. ({\bf b,c}) Time-dependence of the first density moment, i.e., of the center of the mass, $r_{\mathrm{cm}}(t)$ for ({\bf b}) $\Delta=0.2$ and ({\bf c}) $\Delta=0.5$ anisotropy, and various values of exchange decay parameter $\alpha\in\{2.0,2.5,\dots,5.0\}$. Insets of ({\bf b,c}) depict comparison between $r_{\mathrm{cm}}(t)$ and extracted $\overline{\nu}$ for the optimal values of parameter $\alpha$. ({\bf d,e}) Corresponding dependence of the velocity $\nu(t)$ obtained as a time-derivative of $r_{\mathrm{cm}}(t)$ from panels ({\bf c,d}) . Shaded region depict the time window from which average velocity $\bar{\nu}$ is obtained, i.e., $t\in[20,30]$. ({\bf f}) $\alpha$-dependence of $\bar{\nu}$ for various anisotropy $\Delta$. In the same panel we present also the average velocity for the optimal anisotropy $\Delta_{\mathrm{O}}$. Error bars depict the standard deviation of $\overline{\nu}$.  Inset: $\alpha$- and $\Delta$-dependence the averaged velocity $\bar{\nu}$. Points represent the anisotropy $\Delta_{\nu_\mathrm{max}}$ at which we find the largest velocity, while solid line depicts optimal anisotropy $\Delta_\mathrm{O}=\exp(-\alpha+2)$.}
\label{fig1}
\end{figure*}

\section{Spin density expansion}
\label{sec1}

Motivated by the setups studied in the cold-atoms experiments \cite{Ronzheimer2013,Vidmar2013,Neyenhuis2017,Joshi2022}, let us first investigate the expansion of the spin density starting from the nearly down-polarized product state (spin domain)
\begin{equation}
 |\psi(t=0)\rangle=|\underbrace{\uparrow\uparrow\uparrow\uparrow\uparrow}_{x}\underbrace{\downarrow\cdots\downarrow}_{L-x}\rangle\,,
\end{equation}
with $x=5$ up-spins  in a chain with $L=45$ sites (here we assume open boundary conditions). In order to access longer times of spin expansion, the up-sins are initially located at the left  boundary of the studied system. Utilizing the Lanczos time evolution method \cite{PrelBonLan} we monitor the magnetization $\langle S^z_\ell(t) \rangle$ at each site, $\ell$, as exemplified by results presented in Fig.~\ref{fig1}(a). Since the Hamiltonian conserves that total spin projection $S^z_{\rm tot}$, the calculations have been carried out in the subspace with $S^z_{\rm tot}=-35/2$ spanned by $\sim 10^6$ basis states. In order to quantify the dynamics of the system we analyze the velocity $\nu$ obtained from the time derivative of the first density moment, i.e., from the ``center of the mass'', 
\begin{equation}
 \nu(t)=\frac{\mathrm{d}\,r_{\mathrm{cm}}(t)}{\mathrm{d}t}\,,\quad
 r_{\mathrm{cm}}(t)=\frac{\sum_\ell\, \ell \left(\langle S^z_\ell(t)\rangle+1/2\right)}{\sum_\ell\, \left(\langle S^z_\ell(t)\rangle+1/2\right)}\,, \label{vel}
\end{equation}
where the denominator equals $S^z_{\rm tot}/2+L/2$. The detailed $r_{\mathrm{cm}}(t)$ results for exemplary $\Delta=0.2$ and $\Delta=0.5$ and various values of exchange decay parameter $\alpha\in\{2.0,2.5,\dots,5.0\}$ are presented in Fig.~\ref{fig1}(b,c), while the corresponding velocity $\nu(t)$ in Fig.~\ref{fig1}(d,e). Presented results reveal that for given $\Delta$ there exist $\alpha$ for which the $r_\mathrm{cm}(t)$ expands with the largest velocity. 

In order to quantify the behavior of the system for various parameters $(\Delta,\alpha)$, in Fig.~\ref{fig1}(f) we present $\alpha$-dependence of the velocity $\overline{\nu}$ averaged over the time interval $t\in[20,30]$ [shaded region in Fig.~\ref{fig1}(d,e)] for various $\Delta\leq 1$ [see also inset of Fig.~\ref{fig1}(f)]. As evident, $\overline{\nu}$ for fixed $\Delta$ has a nonmonotonic dependence on the decay parameter $\alpha$ with maximum at $\alpha_{\nu_\mathrm{max}}$ (equivalently, for fixed $\alpha$ there exist a maximum at some $\Delta=\Delta_{\nu_\mathrm{max}}$). E.g., for a modest value of the anisotropy $\Delta=0.2$, we find the maximum of $\overline{\nu}$ at $\alpha_{\nu_\mathrm{max}}\simeq3.5$, while $\Delta\to1$ has maximum at $\alpha_{\nu_\mathrm{max}}\to 2$. Furthermore, in the inset of Fig.~\ref{fig1}(f) we present also the position of $\Delta_{\nu_\mathrm{max}}$ as a function of the decay parameter $\alpha$. As one of the main results of this work, we show that the maximum of the velocity at finite times can be found at $\Delta_{\nu_\mathrm{max}} \simeq \Delta_\mathrm{O}=\exp(-\alpha+2)$ (presented also in Fig.~\ref{fig5}). Interestingly, the velocity is approximately constant $\overline{\nu}\simeq J/\sqrt{2}$ along the $\Delta_\mathrm{O}$ line for $\alpha \ge 2$, see black line in Fig~\ref{fig1}(f). Our results indicate that the largest velocity, $\nu_\mathrm{max}\simeq J/\sqrt{2}$, characteristic for the nn free-fermion model \cite{Langer2012,Vidmar2013}, is preserved in the transient dynamics of an interacting chain with the optimal anisotropy $\Delta_\mathrm{O}(\alpha)$.

\begin{figure}[!t]
\includegraphics[width=1.0\columnwidth]{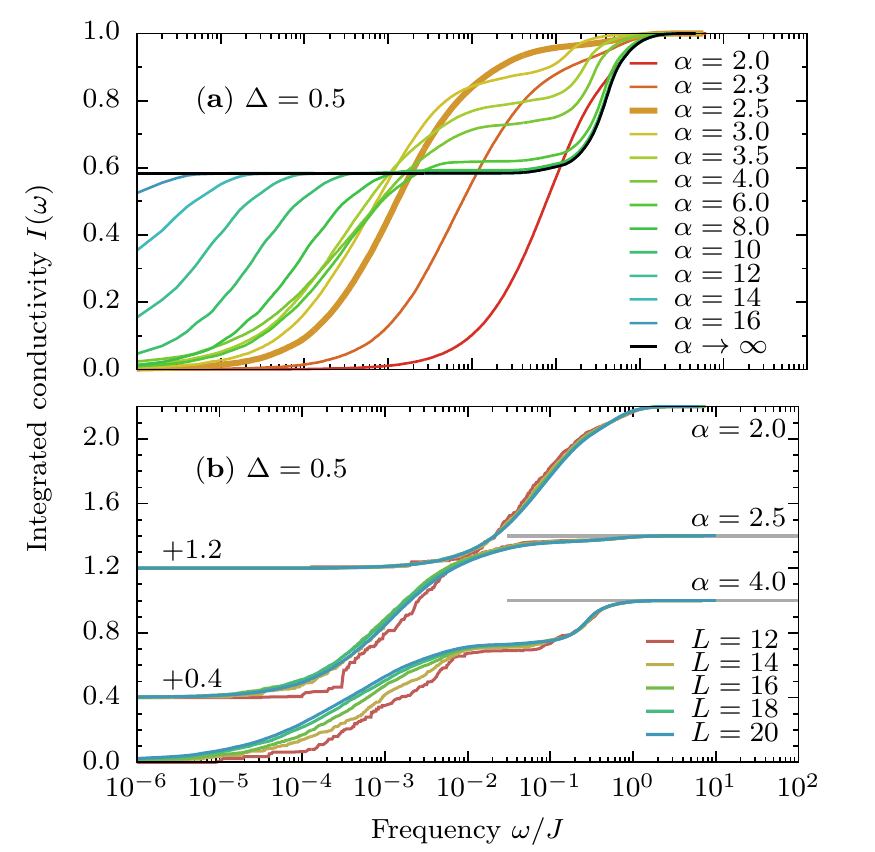}
\caption{({\bf a}) Integrated spin conductivity $I(\omega)$ of long-range AHM, calculated for $L=20$ and $\Delta=0.5$. Thick line represents results for the optimal decay parameter $\alpha_{\sigma}\simeq 2.5 $. ({\bf b}) Size-dependence of $I(\omega)$ for $\Delta=0.5$ below ($\alpha=2.3$), at ($\alpha=2.5$), and above ($\alpha=4.0$) the optimal value $\alpha_{\sigma}$. For clarity, $\alpha=2.5$ and $\alpha=2.0$ results have $0.4$ and $1.2$ offset, respectively.}
\label{fig2}
\end{figure}

\section{Spin conductivity}
\label{sec2}

\begin{figure*}[!htb]
\includegraphics[width=1.0\textwidth]{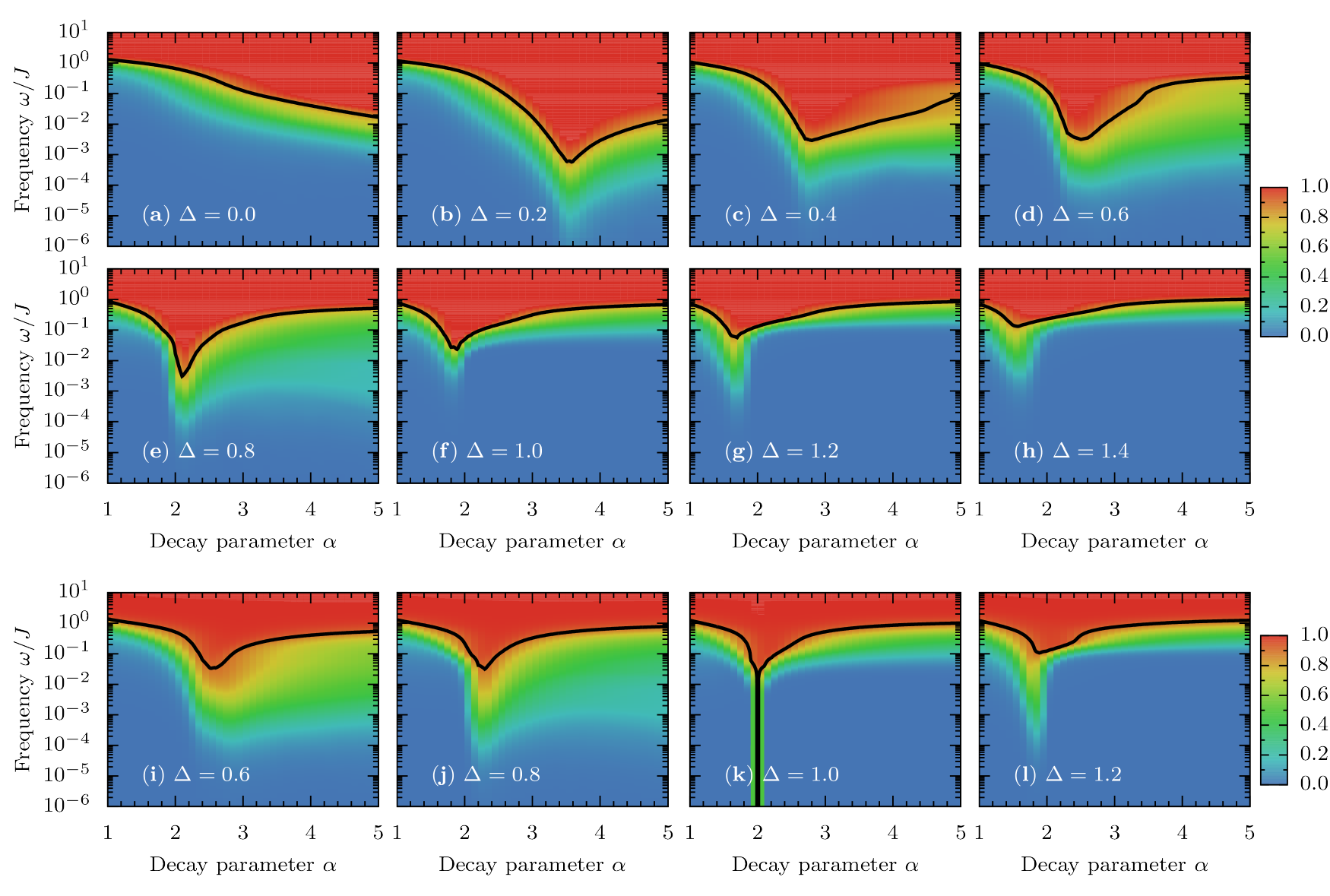}
\caption{({\bf a-h}) $I(\omega)$ in the long-range AHM, for $L=20$ vs. the exchange decay parameter $\alpha$ for various $\Delta$. Note that $\omega$ is shown using the log-scale. Solid lines represent the frequency, $\omega^{*}$, for which $I(\omega^*)=0.9$. ({\bf i-l}) $I(\omega)$ in the generalized HSM for $L=20$. The dip shown by the line in panel ({\bf k}) is consistent with the ballistic transport in the integrable HSM for $\Delta=1$ and $\alpha=2$.}
\label{fig3}
\end{figure*}

In order to gain understanding on the spin dynamics in the long-range AHM, we study the spin conductivity at infinite temperature, see Refs.~\onlinecite{Bertini2021, jaklic} for review. This quantity probes the whole eigenspectrum containing $Z$ states,
\begin{equation}
 \sigma(\omega)=\frac{\pi}{LZ}\sum_{n,m} \langle n|j|m\rangle^2\,\delta(\omega-\epsilon_m+\epsilon_n)\,, \label{opcon}
\end{equation}
where the current operator is obtained from continuity relation, $j=i \sum_{\ell} \ell [H,S^z_\ell]=i/2\sum_{\ell,r} r\,J(r)\left( S^{+}_{\ell}S^{-}_{\ell+r} -\mathrm{H.c.}\right)$ and $H|n\rangle=\epsilon_n |n\rangle$. The results for $\sigma(\omega)$ presented in Fig.~\ref{fig2} and Fig.~\ref{fig3} were obtained with help of exact diagonalization on the system of $L=20$ sites with periodic boundary conditions \cite{PBC} in the largest magnetization sector, i.e., $S^z_{\rm tot}=0$. Furthermore, in order to avoid an arbitrary binning of the finite-size spectra we present the integrated conductivity normalized to the sum rule 
\begin{eqnarray}
 I(\omega) & = &\frac{1}{{\cal S}_{\rm tot}} \int_{-\omega}^{\omega}\mathrm{d}\omega^\prime \sigma(\omega^{\prime})=
 \frac{\pi}{LZ {\cal S}_{\rm tot} } \sum_{n,m} \theta_{mn} \langle n|j|m\rangle^2\,, \nonumber \\
 {\cal S}_{\rm tot} & = & \int_{-\infty}^{\infty}\mathrm{d}\omega^\prime \sigma(\omega^{\prime})=\frac{\pi}{LZ} \sum_{n,m}\langle n|j|m\rangle^2\,,
 \label{sum}
\end{eqnarray} 
so that $I(\omega\to \infty)=1$. Here, $\theta_{mn}=\theta(\omega-|\epsilon_m-\epsilon_n|)$, where $\theta(x)$ is the step function.

In Fig.~\ref{fig2}(a) we show $\alpha$-dependence of $I(\omega)$ for exemplary $\Delta=0.5$. In the $\alpha\to \infty$ limit we recover the standard behavior of the integrable nn-AHM, i.e., the dissipationless Drude $\delta$-peak at $\omega=0$ and a pronounced incoherent (regular) spectrum at $\omega/J \sim 0.1 - 1$ \cite{Prelovsek2021}. Upon increasing the range of spin exchange ($\alpha$ decreases down to $\alpha\simeq 10$) we observe broadening of the Drude $\delta$-peak, with no effect on the incoherent part of the spectrum. Such behavior is consistent with the integrability-breaking of the nn-AHM. However, for $\alpha<10$, the incoherent part of $\sigma(\omega)$ shifts towards lower frequencies. E.g., for $\alpha = 2.5$ the most of the spectral weight is contained below $\omega/J<0.01$, i.e., $I(\omega/J > 0.01 ) \simeq 1$. It means that up to extremely long times, $t \sim 100/J$, the spin dynamics in the studied systems is the same as the ballistic particle dynamics in a system of noninteracting particles where $\sigma(\omega) \propto \delta(\omega)$ so that $I(\omega > 0 ) = 1$. Surprisingly, further reduction of $\alpha$ reverses this behavior and the frequency range of the incoherent part increases.

\begin{figure}[!hb]
\includegraphics[width=1.0\columnwidth]{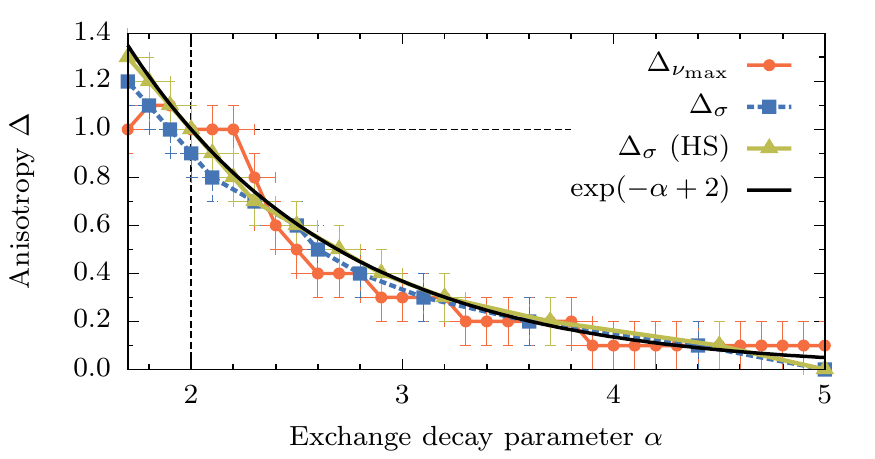}
\caption{Exchange decay $\alpha$ dependence of the optimal anisotropy $\Delta$ as extracted from analysis of the spin domain expansion (discussed in Sec.~\ref{fig1}), i.e., $\Delta_{\nu_\mathrm{max}}$, and from behaviour of optical conductivity, i.e., $\Delta_\sigma$. Black solid line depicts $\Delta_\mathrm{O}(\alpha)=\exp(-\alpha+2)$ prediction. Error bars depict accuracy of the data grid on base of which given curve was obtained.}
\label{fig4}
\end{figure}

In order to quantify this behavior, in Fig.~\ref{fig3}(a-h) we present the frequency-dependence of $I(\omega)$ for various $\alpha$ and $\Delta$. The solid lines mark the frequencies, $\omega^*$, for which $I(\omega^*)$ covers $90\%$ of the total sum rule. For appropriately tuned parameters $\alpha$ and $\Delta$, $\omega^*$ becomes unexpectedly small, $\sim 10^{-3} - 10^{-2} $, hence the studied systems exhibit a transient ballistic transport up to very long times $t^{*}\sim J/\omega^{*}$ (consistent with the results discussed in Sec.~\ref{sec1}). We define the optimal $\Delta_{\sigma}$ as a value of anisotropy when $\omega^*$ is the smallest for fixed $\alpha$ or, equivalently, the optimal $\alpha_{\sigma}$ for fixed $\Delta$. The latter quantity can be identified via the minima of the black curves presented in Fig.~\ref{fig3}. Figure~\ref{fig4} shows that $\Delta_{\sigma}$ obtained from low-frequency dynamics of a system with periodic boundary conditions is in perfect agreement with $\Delta_\mathrm{O}(\alpha)$ established previously from the short-time expansion of the spin domains in a chain with open boundary conditions. The latter clearly shows that the presence of the optimal anisotropy does not emerge as a finite-size effect that might have been expected in a finite-size chains with long-range interactions (the issue to which we will come back in Sec.~\ref{sec4}). Furthermore, our finite-size analysis, presented in Fig.~\ref{fig2}(b) indicates that the size-dependence is rather inessential for $\alpha \sim 2$ whereas for larger $\alpha$, the incoherent part of $ \sigma(\omega)$ is shifted with increasing $L$ to even smaller frequencies.

Interestingly, the exponential dependence, \mbox{$\Delta_\mathrm{O} = \exp(-\alpha+2)$}, extends also to the regime with $\alpha<2$ when $\Delta_{\sigma}>1$, see Fig.~\ref{fig3}(g) and \ref{fig3}(h), whereas the ballistic transport in the integrable nn-AHM is limited to $\Delta <1$ \cite{Znidaric2011,Herbrych2011,Bertini2021}. Note however that our available system sizes (i.e., $L=20$ for ED method) don't allow for accurate investigation of $\alpha<2$ region.

\subsection*{Haldane-Shastry-like model}
\label{sec3}

Related to our investigations is the integrable Haldane-Shastry model (HSM) \cite{Haldane1988,Shastry1988} where $\Delta=1$ and \mbox{$J(r)=\sin^2(\pi/L)/\sin^2(\pi r/L)$}. Within this model, the spin current operator $j$ is fully conserved in $L \to \infty$ limit \cite{Sirker2011,Defenu2021}, and one obtains a purely ballistic transport as it is the case for noninteracting particles [$\alpha\to\infty$, $\Delta=0$ of \eqref{ham}]. Note also that in the thermodynamic limit, the spin exchange in HSM is identical to our investigations, since $\sin^{-2}(r/L) \propto r^{-2}$ for $r\ll L$. However, for any finite $L$ (especially, for numerically accessible $L\simeq20$), $J(r) \propto 1/r^2$ breaks the integrability of underling model.

Here, we generalize the HSM model allowing for a finite spin anisotropy $\Delta\ne1$ and $\alpha$-dependent exchange interaction, i.e., we use $J(r)=J_\mathrm{HS}(r)$ in Eq.~\eqref{ham} with
\begin{equation}
 J_\mathrm{HS}(r)=J \frac{\sin^\alpha(\pi/L)}{\sin^\alpha(\pi\,r/L)}\,.
\end{equation}
Figure~\ref{fig3}(i-l) shows the integrated optical conductivity for the generalized HSM, which is very similar to the conductivity in the long-range AHM with $J(r)\propto 1/r^{\alpha}$. In particular, we observe that the HSM reproduces all main features of $I(\omega)$ which were obtained for AHM and interpreted in terms of the quasiballistic transport in the latter model. The only exception concerns the point $\Delta=1$ and $\alpha=2$ for which we recover the integrability of HSM with the Drude-peak at $\omega=0$. The latter peak shows up as a dip in the solid line in Fig.~\ref{fig3}(k). For the studied system with $L=20$, the Drude contribution at the integrable point contains {\bf $94\%$} of the total sum-rule, whereas it should contain the entire spectral weight in the $L\to\infty$ limit. Note that in the generalized HSM the specific form of the anisotropy which corresponds to the optimal quasiballistic transport, $\Delta_\mathrm{O}(\alpha)=\exp(-\alpha+2)$, smoothly interpolates between two models where the transport is purely ballistic: the free-particle model for $\Delta=0$, $\alpha \to \infty$ and the integrable HSM for $\Delta=1$ and $\alpha=2$. Since the spin conductivity is almost identical between the $1/\sin(\pi r/L)^{\alpha}$ and $1/r^\alpha$ models, they share the same set of parameters which are optimal for the ballistic transport.

\begin{figure*}[!htb]
\includegraphics[width=0.975\textwidth]{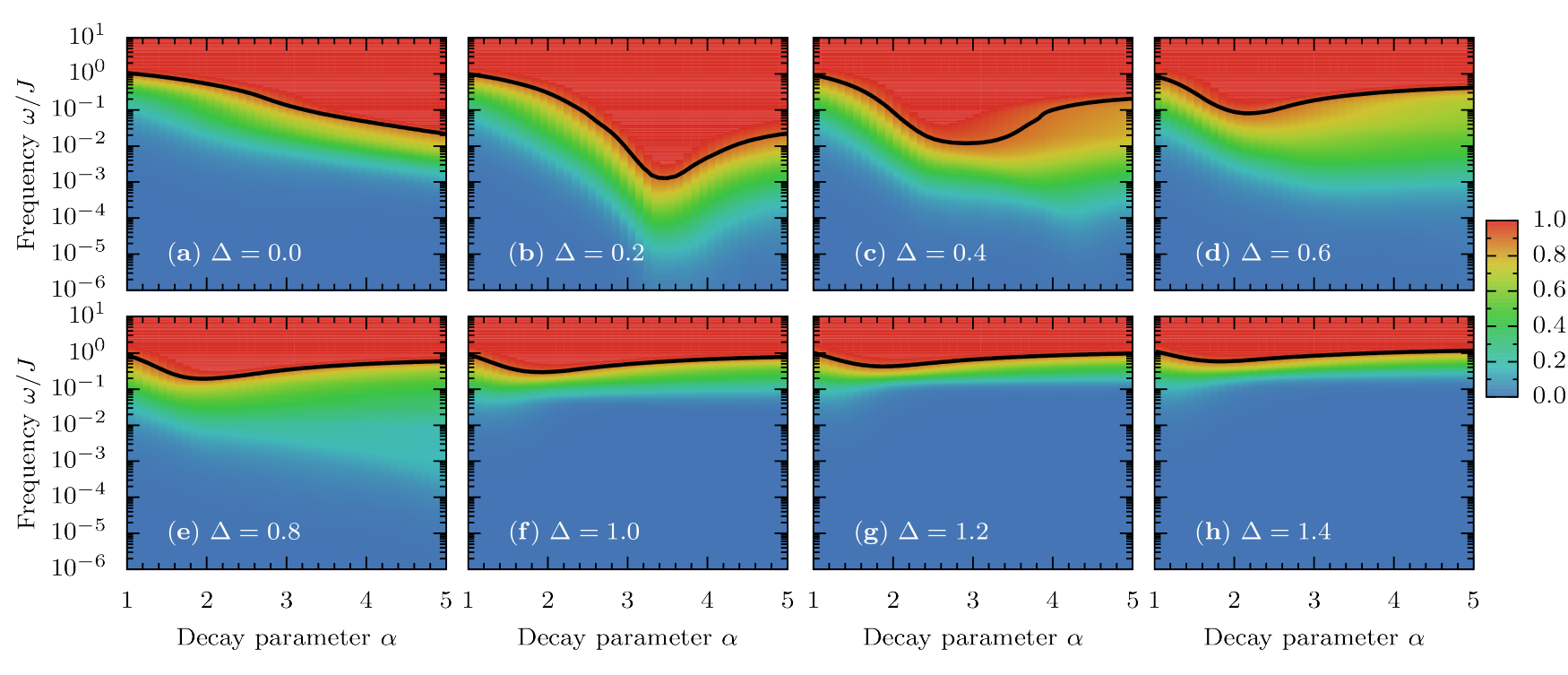}
\includegraphics[width=0.975\textwidth]{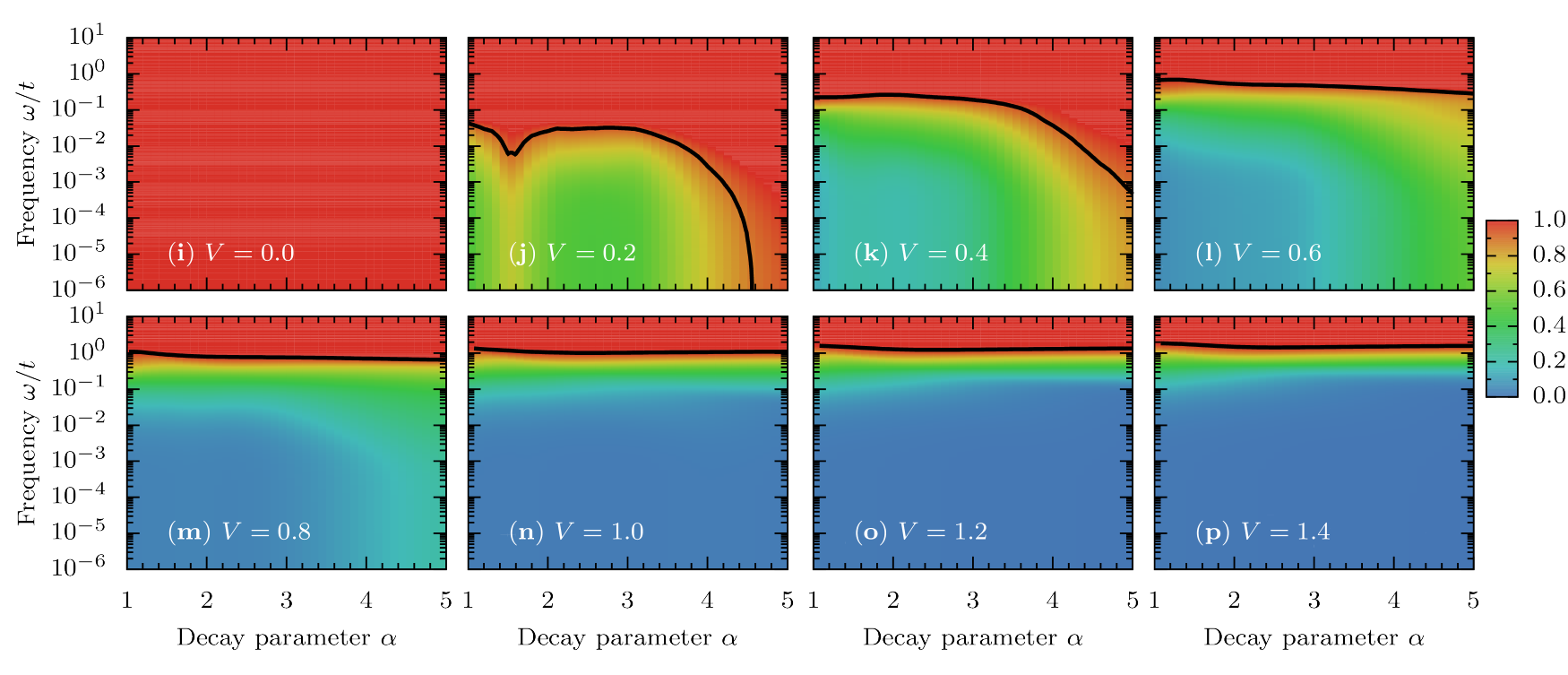}
\caption{$I(\omega)$ in ({\bf a}-{\bf h}) the $\alpha$-dependent $J_1$-$J_2$ AHM model, Eq.~\eqref{hamj1j2}, and in ({\bf i}-{\bf p}) the long-range $t-V$ model of spinless fermions, Eq.~\eqref{hamtv}, for $L=18$ vs. the exchange decay parameter $\alpha$. Note that $\omega$ is shown using the log-scale. Solid lines represent the frequency, $\omega^{*}$, for which $I(\omega^*)=0.9$.}
\label{fig8}
\end{figure*}

\subsection*{Spin conductivity of related models}
In Fig.~\ref{fig3} we have shown the detailed frequency, $\omega$, dependence of the spin conductivity $\sigma(\omega)$ for the anisotropic Heisenberg model with long-range exchange $J(r)=1/r^\alpha$ and for the generalized Haldane-Shastry model (HSM) with $J(r) \propto 1/ \sin^\alpha(\pi\,r/L)$. To verify that the phenomenon presented is unique to the long-range AHM, we have investigated the behavior of $\sigma(\omega)$ in two other models relevant for our analysis. First, we have considered the $J_1-J_2$ AHM, i.e., the model \eqref{ham} limited to the nearest- ($r=1$) and next-to-nearest ($r=2$) neighbor spin exchange,
\begin{align}
  H_{J_1-J_2}=& \sum_{\ell} J\left[ \frac{1}{2}\left(S^{+}_{\ell}S^{-}_{\ell+1}
 +S^{-}_{\ell}S^{+}_{\ell+1}\right)+\Delta S^{z}_{\ell}S^{z}_{\ell+1}\right]\nonumber \\
 +&\sum_{\ell} \frac{J}{2^\alpha}\left[ \frac{1}{2}\left(S^{+}_{\ell}S^{-}_{\ell+2}
 +S^{-}_{\ell}S^{+}_{\ell+2}\right)+\Delta S^{z}_{\ell}S^{z}_{\ell+2}\right]\,.
\label{hamj1j2}
\end{align}
Results presented in Fig.~\ref{fig8}(a-b) for $\Delta=0.0-0.2$ are akin to the full long-range AHM. Such behavior is expected since for small value of anisotropy $\Delta$, the optimal decay parameter is $\alpha\sim4$ and, consequently, $r\gtrsim 3$ terms in \eqref{ham} have small overall contribution. Further increasing $\Delta$ yields rather featureless conductivity $\sigma(\omega)$ with most of the spectral weight at $\omega/J\sim0.1$. As a consequence we conclude that the long-range exchange $J(r)$ is necessary for the quasiballistic transport at extremely small $\omega$-scales. Second, we have examined the fermionic version of the model, i.e., the long-range $t-V$ model of interacting spinless fermions,
\begin{align}
 H_{t-V}=\sum_{\ell,r} \frac{1}{r^\alpha} &\bigg[\frac{t}{2}\left(c^{\dagger}_{\ell}c^{\phantom{\dagger}}_{\ell+r}
 +c^{\phantom{\dagger}}_{\ell}c^{\dagger}_{\ell+r}\right)\nonumber \\
 &+V(n_{\ell}-1/2)(n_{\ell+r}-1/2)\bigg]\,,
  \label{hamtv}
\end{align}
with $t=1$ and $V$ as the strength of interaction. Here, $c^{\left(\dagger\right) }_{\ell}$ obey the fermionic commutation relations. Note that in the $\alpha\to\infty$ limit the above model and long-range AHM~\eqref{ham}  are identical, i.e. they are related to each via the Jordan-Wigner transformation. Our results reveal that the quasiballistic transport is absent in this model. As a test, we confirm in Fig.~\ref{fig8}(i) that the free particles ($V=0$) are always ballistic [with $\sigma(\omega)=D\delta(\omega)$] for all values of the decay parameter $\alpha$. On the other hand, for  finite interaction  $0<V<1$ our results show that in the fermionic model, the optimal behavior occurs in the $\alpha\to\infty$ limit, i.e., for the  nearest-neighbor $t-V$ model (for nn-AHM) where the ballistic transport of particles originates from the integrability.  We recall  that for $V>1$, the transport in the integrable nn-AHM chain ($\alpha \to \infty$) is diffusive \cite{Steinigeweg2009,Karrasch2014,Steinigeweg2015,Prelovsek2022}. Figs. \ref{fig8}(o) and \ref{fig8}(p) suggest that for $V>1$, the conductivity in the long-range $t$-$V$ model weakly depends on $\alpha$. Therefore, we expect that for $V>1$ the transport is diffusive for all studied $\alpha >1$.

\begin{figure}[!t]
\includegraphics[width=1.0\columnwidth]{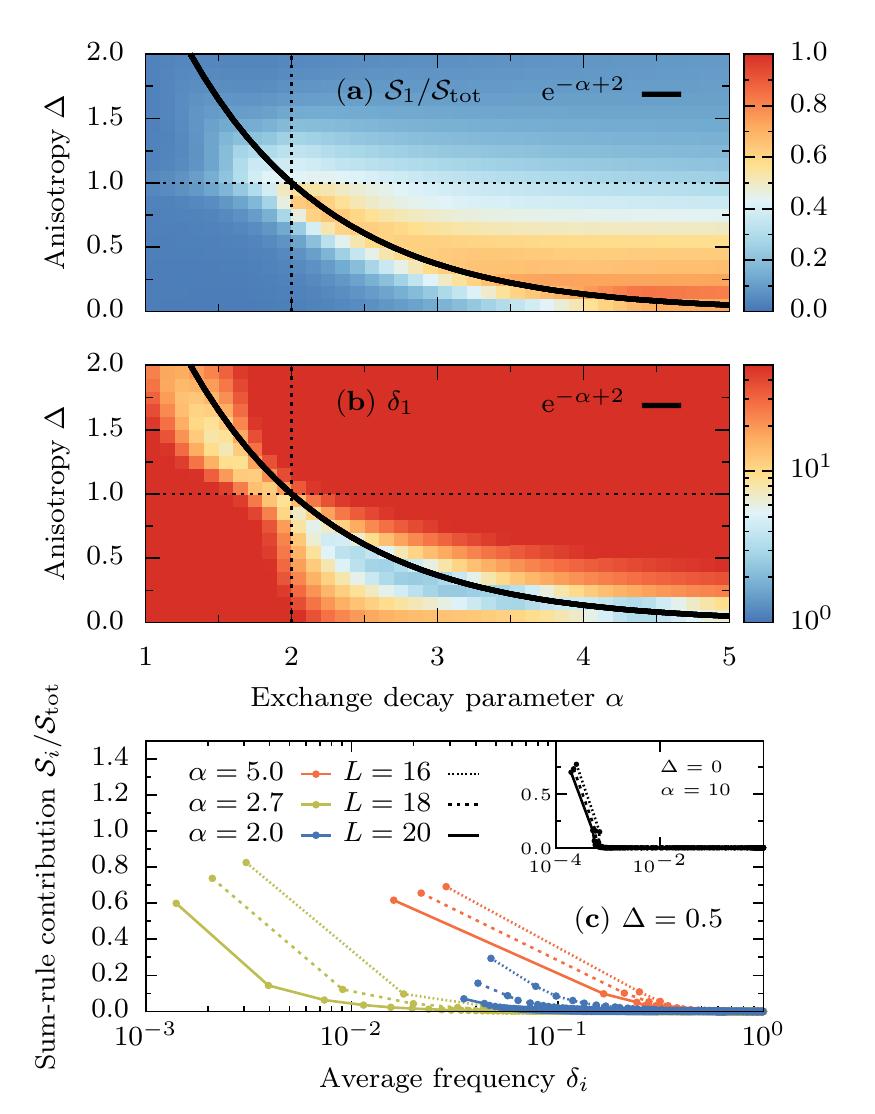}
\caption{({\bf a}) Leading contribution to the spectral sum of spin conductivity, ${\cal S}_1$, accounting for $Z$ (out of total $Z^2$) matrix elements. ${\cal S}_1$ is obtained from $L=20$ and normalized to the total spectral sum ${\cal S}_{\rm tot}$. ({\bf b}) Color background shows the average frequency $\delta_1$ of dominating contribution ${\cal S}_1$ to $\sigma(\omega)$. Here, the values of $\delta_1$ are normalized to the average level spacing. In both panels the optimal anisotropy $\Delta_\mathrm{O}=\exp(-\alpha+2)$ is also presented. ({\bf c}) Partial contributions to the spectral sum, ${\cal S}_1$, vs. the corresponding frequencies $\delta_i$. Points connected by each line correspond to ${\cal S}_1$, ${\cal S}_2$,...,${\cal S}_{10}$ ordered from left to right. Here, ($\Delta=0.5\,,\alpha=2.7$) corresponds to the optimal case.}
\label{fig5}
\end{figure}

\section{Similarity to noninteracting fermions}
\label{sec4}

Spin conductivity in the integrable nn-AHM with \mbox{$\Delta \lesssim 1$} contains significant regular part that extends over a broad range of frequencies \cite{Prelovsek2021}. On the other hand, the regular part is absent in the nn-AHM for $\Delta= 0$ (noninteracting fermions) as well as in the thermodynamic limit of the HSM \cite{Sirker2011}. In the case of the long-range AHM with optimal $\Delta_{\sigma}$, the regular part of $\sigma(\omega)$ is accumulated at very small albeit nonzero frequencies, see Fig.~\ref{fig3}.  Consequently a decay of the spin current in long-range spin model  corresponds to a time-scale that is much larger than the times-scale for the (partial) decay of the current in the nn-AHM. Therefore the spin transport in the long-range AHM with $\Delta_{\mathrm{O}}(\alpha)$ resembles the ballistic transport of noninteracting particles rather then a more generic case of nn-AHM model. In oder to demonstrate this similarity in more detail, in this Section we show that for each eigenstate of the Hamiltonian $| n \rangle$ there exist only a few matrix elements $ \langle m | j | n \rangle$ which significantly contribute to the optical spin conductivity and to the corresponding sum rule, ${\cal S}_{\rm tot}$, in Eq.~\eqref{sum}. 

In the long-range AHM with $\Delta=\Delta_\mathrm{O}$ the current operator $j$ has large matrix elements only for states close in energy. This behavior resembles the case of purely ballistic systems where $[H,j]=0$ and the latter states have equal energies. For each $|n\rangle$ we sort the eigenstates $\{ | m \rangle \}$ in descending order of $\langle n |j|m \rangle^2 $ and denote the sorted states as $|n_i\rangle,\; i=1,\dots,Z$. The spectral sum can be represented as the sum of partial contributions ${\cal S}_{\rm tot}={\cal S}_1+{\cal S}_2+\dots+{\cal S}_Z$ where ${\cal S}_i=\pi(L Z)^{-1}\sum_n \langle n |j|n_i \rangle ^2$ and ${\cal S}_1 \ge {\cal S}_2 \dots \ge{\cal S}_Z $. We determine also the (weighted) average frequencies of these contributions, cf. Eq.~\eqref{opcon},
\begin{equation}
 \delta_i=\frac{\sum_n \langle n |j|n_i \rangle ^2 \;|\epsilon_n-\epsilon_{n_i} | }{ \sum_n \langle n |j|n_i\rangle ^2 }\,.
\end{equation}
Figure~\ref{fig5}(a) depicts the ratio ${\cal S}_1/{\cal S}_{\rm tot}$ whereas the color background in Fig.~\ref{fig5}(b) shows the frequency $\delta_1$ normalized to the average level spacing. For the optimal anisotropy $\Delta_\mathrm{O}$ one observes that  ${\cal S}_1$ covers a majority of the  spectral sum, see bright colors in fig.~\ref{fig5}(a). Results in fig.~\ref{fig5}(b) demonstrate that ${\cal S}_1$ originates from the matrix elements $\langle n | j| m\rangle$  between eigenstates $|n \rangle$ and $|m \rangle$ for which the energy difference $|\epsilon_n-\epsilon_m|$ is of the order of several level spacings (for  accessible $L$).  It is not the case for other $\Delta$, when ${\cal S}_1$ is significantly smaller and the corresponding frequency $\delta_1$ is significantly larger than for the optimal $\Delta_{\rm O}$. In order to visualize the finite-size effects, in  Fig.~\ref{fig5}(c) we show the ratio ${\cal S}_i/{\cal S}_{\rm tot}$ vs. $\delta_i$ for various $L$. Points on each line correspond to ${\cal S}_1$, ${\cal S}_2$,...,${\cal S}_{10}$ which are ordered from left to right. Upon increasing $L$, the contributions to $\sigma(\omega)$ which come from the matrix elements entering ${\cal S}_{i>1}$ becomes more significant also for the optimal anisotropy. However, we observe that $\delta_i$ remains small whenever ${\cal S}_i$ is large. Therefore for the optimal anisotropy, the regular part of the optical conductivity is confined to the low frequency regime for all accessible $L$. The same message arises also from results in Fig.~\ref{fig2}(b).

\section{Slowly-decaying fermionic currents}
\label{sec5}

In previous Sections we have shown that at the optimal anisotropy $\Delta_\mathrm{O} = \exp(-\alpha+2)$ the spin current $j\propto \sum_{\ell,r}(S^+_{\ell}S^-_{\ell+r}-\mathrm{H.c.})$ is decaying slowly. Here we will demonstrate that on the $\Delta_\mathrm{O}(\alpha)$ line one can find a broad class of physically relevant observables which display similar dynamics. In order to identify such operators, we apply a numerical algorithm that was originally applied for identifying an orthogonal set of local (and quasilocal) integrals of motion \cite{Mierzejewski2015-1,Mierzejewski2015-2}. Except for the known integrable points, $\alpha \to \infty$ and $\Delta=1$ with $\alpha=2$ for HSM, we do not expect the presence of local integrals of motion in the thermodynamic limit. Nevertheless, one can expect that found operators will exhibit finite stiffness, i.e., nondecaying component in finite-size system, which should indicate slow dynamics in macroscopic chains. 
 
For the sake of completeness, we recall the main steps of the numerical algorithm. First, we construct all translationally-invariant current-like operators supported on up to $M$ consecutive lattice sites
\begin{equation}
 A_{\gamma}= i \frac{1}{\sqrt{L}}  \sum_{\ell}\left(  o_{\ell+1} o_{\ell+2} ... o_{\ell+M} -{\rm H.c.}\right) \,,\label{epr1}
\end{equation}
where $o_{\ell} \in \{ 1_{\ell},\sqrt{2} S^{-}_{\ell}, \sqrt{2} S^{+}_{\ell},2 S^z_{\ell} \}$ and the index $\gamma$ enumerates the resulting operators. We then expressed all $A_{\gamma}$ in the basis of the eigenstates of the Hamiltonian, $H |n \rangle=E_n |n \rangle $. The conserved part of $A_{\gamma}$ is determined by the matrix elements between the degenerate eigenstates, 
\begin{equation}
 \bar{A}_{\gamma}=\sum_{m,n:\;E_m=E_n} \langle m|A_{\gamma}|n\rangle | m \rangle \langle n|, \quad \quad [\bar{A}_{\gamma},H]=0\,.
\end{equation} 
The final step is the numerical diagonalization of a matrix that contains averaged products of the latter operators, $M_{\gamma \gamma' }= \langle \bar{A}_{\gamma} \bar{A}_{\gamma'}\rangle= (1/Z){\rm Tr}(\bar{A}_{\gamma} \bar{A}_{\gamma'} ) $, where the trace is carried out over all many-body states in all $S^z_{\rm tot}$ sectors. Namely, we numerically solve the eigenproblem
\begin{equation}
 \sum_{\gamma,\gamma'} (O^{T})_{a \gamma} M_{\gamma \gamma' } O_{\gamma' a'} = \lambda_a \delta_{a,a'}\,, 
\end{equation}
where $O$ is an orthogonal matrix.

\begin{figure*}[!ht]
\includegraphics[width=1.0\textwidth]{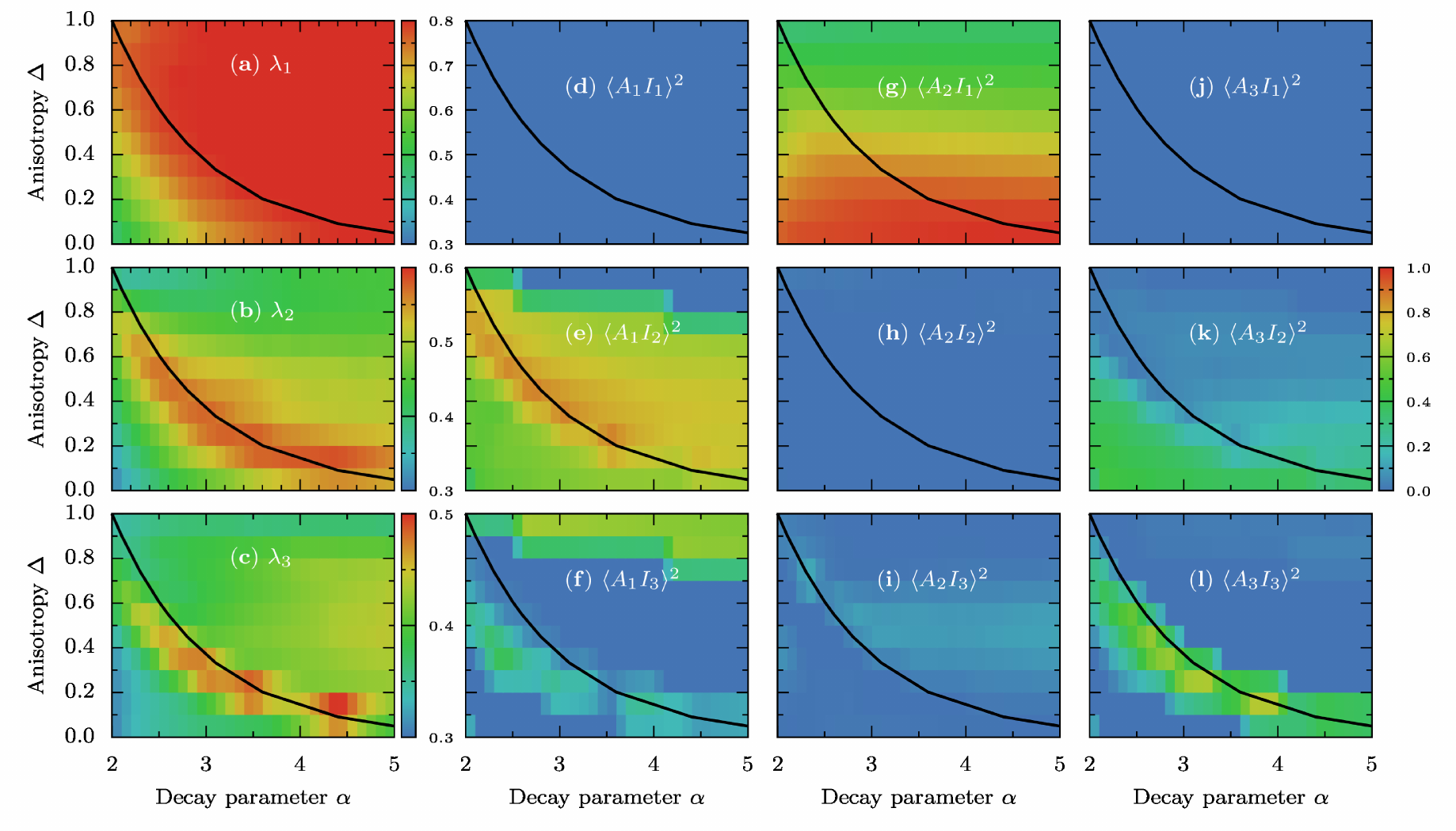}
\caption{({\bf a}-{\bf c}) The eigenvalues (stiffnesses) $\lambda_a$ of three local operators $I_a$ with the largest conserved part. The remaining panels ({\bf d}-{\bf l}) depict projections of $I_a$ on the local operators defined in Eq.~\eqref{currenty}, $ \langle A_{\gamma} I_{a} \rangle^2= |O_{\gamma a} |^2$ for $a=1,2,3$ and $\gamma=1,2,3$. See text for details.}
\label{fig6}
\end{figure*} 

The eigenvectors of $M$ define a sequence of local operators, $I_a=\sum_{\gamma} O_{\gamma a} A_{\gamma}$, which are mutually orthogonal, $\langle I_{a} I_{a'} \rangle =\delta_{a,a'}$, and the squared norms of their conserved parts (called stiffnesses from now on) are equal the corresponding eigenvalues, $||\bar{I}_a ||^2 = \langle \bar{I}_a \bar{I}_a \rangle =\lambda_a$. All stiffnesses are bounded $1\ge \lambda_1\ge \lambda_2 \ge \dots \ge 0$ while local integrals of motion correspond to $\lambda_a=1$. Again, we want to point out that one does not expect the latter for nonintegrable $J(r)=1/r^\alpha$ system studied here (except in $\alpha\to\infty$ limit). 

Figure~\ref{fig6} depicts the first three largest eigenvalues $\lambda_a$ and corresponding operators $I_a$, obtained for $M=4$ sites support. In Fig.~\ref{fig6}(a-c) we present the value of stiffnesses itself obtained for various $\Delta$ and $\alpha$. It is evident from the presented results that second and third eigenvalue, $\lambda_2$ and $\lambda_3$, respectively, have the largest value at $\Delta_\mathrm{O}$. On the other hand, $\lambda_1$ does not have clear maximum at $\Delta_\mathrm{O}$, but rather constant value for $\Delta(\alpha)>\Delta_\mathrm{O}$ and sharp drop for $\Delta(\alpha)<\Delta_\mathrm{O}$.

\begin{figure}[!htb]
\includegraphics[width=1.0\columnwidth]{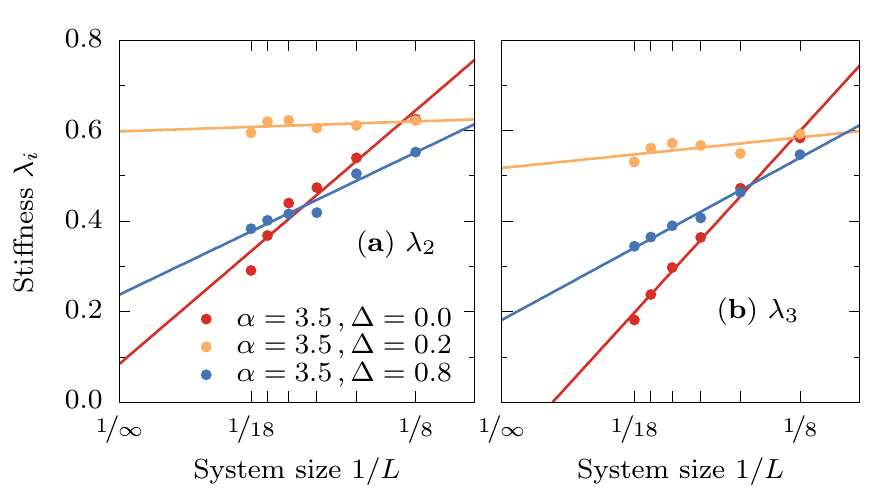}
\caption{Finite-size dependence of (a) second $\lambda_2$ and (b) third $\lambda_3$ eigenvalue (stiffness) at optimal anisotropy ($\Delta=0.2\,,\alpha=3.5$) and away from this point ($\Delta=0.0$ and $\Delta=0.8$).}
\label{fig7}
\end{figure}

Figure~\ref{fig7} shows the finite-size dependence of the stiffness $\lambda_2$ and $\lambda_3$ for a set of parameters corresponding to the optimal anisotropy, $\Delta=3.5, \alpha=0.2$ as well as for parameters which are shifted away from it. One may observe that not only the stiffnesses are maximal for $\Delta=\Delta_\mathrm{O}$, but also the finite-size effect are least visible. We are not aware of any mechanism that might explain the presence of nonzero stiffnesses in the studied chains in the thermodynamic limit. For this reason we expect that the stiffnesses vanish also for the optimal anisotropy, however their decay becomes visible for much larger systems than for other model parameters.

Finally, our analysis of the most conserved operators $I_a$ [corresponding to the eigenvalues in Fig.~\ref{fig6}(a-c)] indicated that the largest contributions come from the local current-like operators of the form:
\begin{eqnarray}
A_{1}& = & i \frac{2}{\sqrt{L}} \sum_{\ell} \left(S^{+}_{\ell} S^{-}_{\ell+1} -{\rm H.c.}\right)\,,\nonumber \\
A_{2}& = & i \frac{4}{\sqrt{L}} \sum_{\ell}\left(S^{+}_{\ell} S^z_{\ell+1} S^{-}_{\ell+2} -{\rm H.c.}\right)\,,\nonumber \\
A_{3}& = & i \frac{8}{\sqrt{L}} \sum_{\ell} \left(S^{+}_{\ell} S^z_{\ell+1} S^z_{\ell+2}S^{-}_{\ell+3} -{\rm H.c.}\right)\,.
\label{currenty}
\end{eqnarray}
The latter contributions are formally quantified via projections $ \langle A_{\gamma} I_{a} \rangle^2= |O_{\gamma a} |^2$ and are shown in Fig.~\ref{fig6}(d-l). Since $\sum _{\gamma} |O_{\gamma a} |^2 =1$, the most conserved operator, $I_1$, has the largest projection on the current $A_2$ which is even under the spin-flip transformation, whereas $I_2$ and $I_3$ have the largest projections on $A_1$ and $A_3$, respectively. The latter two operators are odd under the spin-flip transformation, which maps \(S^z_i\) to \(-S^z_i\), \(S^+_i\) to \(S^-_i\) and \(S^-_i\) to \(S^+_i\). The Wigner-Jordan transformation maps the spin-current operators Eq.~\eqref{currenty} to fermionic currents which describe hoppings between the first, the second and the third nearest-neighbors. Despite the studied spin-system can not be mapped on a fermionic model with two-body interactions, numerical results discussed in the present Section consistently support the presence of slowly decaying odd fermionic currents which are most stable at $\Delta_\mathrm{O}(\alpha)$. We also note that one of the studied local currents, $A_1$, has a large projection on the actual spin current that was studied in Sec.~\ref{sec2}.

\section{Conclusion}
\label{sec6}

We have studied the spin transport in a Heisenberg chain with long-range exchange $J(r)= J/r^{\alpha}$ and spin anisotropy $\Delta$. We have found that the system may be tuned to a quasiballistic regime, where the spectral sum of the spin conductivity is accumulated at frequencies smaller than $\omega^*/J\sim 10^{-3} - 10^{-2}$. Such systems exhibits a ballistic spin transport that is transient but persists up to unexpectedly long times $t\sim 1/\omega^*$, consistent with our domain wall expansion analysis. The quasiballistic transport occurs not just in the neighborhood of known ballistic systems but rather along a sharp line in the space of model parameters, $\Delta \simeq \exp(-\alpha+2)$. The latter line smoothly interpolates between two models in which the transport is ballistic, i.e., the regular part of conductivity is absent. Namely, the line interpolates between free particles for $\alpha=\infty$, $\Delta=0$ and the Haldane-Shastry-like model for $\alpha=2$, $\Delta=1$. Furthermore, the presence of a long-living quasiballistic transport resembles the phenomenon of prethermalization \cite{Berges2004,Moeckel2008,Kollar2011} that is present in nearly-integrable models. Indeed, our results may suggest that the studied Heisenberg chain could be viewed as nearly-integrable for all parameters satisfying the relation $\Delta \simeq \exp(-\alpha+2)$, and not in just small parameter space close to the known integrable points. Our numerical studies indicate that the quasiballistic dynamics can be observed not only from the spin current but  also from a broader class of fermionic-like operators, despite the studied spin chain can not be mapped on a fermionic model with two-body interaction. 

In the case of large $\alpha \gg 1$ we have found that the optimal anisotropy $\Delta \simeq \exp(-\alpha+2) \ll1 $ can be identified also in the spin chain that includes first nearest neighbor ($J$) and second nearest neighbor ($J_2=J/2^{\alpha}$) terms. However, the long-lasting ballistic transport discussed above is specific to the spin chains and does not occur in a fermionic system with long-range hoppings and interactions. The origin of the optimal anisotropy and its exponential dependence on $\alpha$ remain  open problems. A possible (but speculative) scenario is that the studied spin chain with $\Delta \simeq \exp(-\alpha+2)$ is a good approximation to a more complex spin model that is integrable and conserves the spin current. Then, the values $\omega^*$ would be determined by the quality of this approximation.

\begin{acknowledgments}
The authors thank R. Steinigeweg and P. Prelov\v{s}ek for fruitful discussions. M. Mierzejewski, J. Wronowicz and J. Paw\l owski acknowledge support by the National Science Centre (NCN), Poland via project 2020/37/B/ST3/00020. J.~Herbrych acknowledges support by the Polish National Agency for Academic Exchange (NAWA) under contract PPN/PPO/2018/1/00035 and by the National Science Centre (NCN), Poland via project 2019/35/B/ST3/01207. The calculations were carried out using resources provided by the Wroc{\l}aw Centre for Networking and Supercomputing.
\end{acknowledgments}



\begin{thebibliography}{99}
\bibitem{PBC} In order to avoid the double counting of $J(r)$ on the system with periodic boundary condtions we have restricted the sum over $r$ in \eqref{ham} to $L/2-1$.
\end{thebibliography}
\end{document}